\documentclass[a4paper,twocolumn]{esapub2005} 

\usepackage{times,natbib,graphicx}

\newcommand{\mic}{\mbox{$\mu$m}}

\newcommand{\degg}{$^\circ$}

\newcommand{\pccm}{\mbox{cm$^{3}$}}

    \def\independenT#1#2{\mathrel{\setbox0\hbox{$#1#2$}%
    \copy0\kern-\wd0\mkern4mu\box0}} 
    
\newcommand{\Lower}[1]{\smash{\lower 1.5ex \hbox{#1}}}

\newlength{\captsize}   \let\captsize=\footnotesize
\newlength{\captwidth}  
\newlength{\beforetableskip} 
\newcommand{\capt}[1]{\begin{minipage}{\captwidth}
\let\normalsize=\captsize
\caption[#1]{#1}
\end{minipage}\\ \vspace{\beforetableskip}}

\makeatletter

\begin{document}

\pagestyle{empty}

\title{GEO Debris and Interplanetary Dust: Fluxes and Charging Behavior}
\author{A. L. Graps}
\affil{INAF Istituto di Fisica dello Spazio Interplanetario, CNR-ARTOV,
Via del Fosso del Cavaliere 100, 00133 Rome, Italy, Amara.Graps@ifsi-roma.inaf.it}
\author{Green, S.F.}
\author{McBride, N.}
\affil{PSSRI, The Open University, Walton Hall, Milton Keynes MK7 6AA, U.K. }
\author{McDonnell, J.A.M.}
\affil{Unispace Kent, P.O. Box 318, Canterbury, Kent CT2 8HB, U.K.}
\author{Bunte, K.}
\affil{eta\_max space GmbH, Richard-Wagner-Strasse 1, D-38106 Braunschweig, Germany}
\author{Svedhem, H.}
\author{Drolshagen, G.}
\affil{ESA/ESTEC, PB 299, NL-2200 AG Noordwijk, The Netherlands}



\maketitle

\begin{abstract}

In September 1996, a dust/debris detector: GORID was launched into 
the geostationary (GEO) region as a piggyback instrument on the
Russian Express-2 telecommunications spacecraft. The instrument began
its normal operation in April 1997 and ended its mission in July 2002.
The goal of this work was to use GORID's particle data to identify and
separate the space debris to interplanetary dust particles
(IDPs) in GEO, to more finely determine the instrument's measurement
characteristics and to derive impact fluxes. While
the physical characteristics of the GORID impacts alone are
insufficient for a reliable distinction between debris and
interplanetary dust, the temporal behavior of the impacts are strong
enough indicators to separate the populations based on clustering.
Non-cluster events are predominantly interplanetary, while cluster
events are debris. The GORID mean flux distributions 
(at mass thresholds which are impact speed dependent) for IDPs,
corrected for dead time, are 1.35~$\times$~10$^{-4}$~m$^{-2}$s$^{-1}$
using a mean detection rate: 0.54~d$^{-1}$, and for space debris are
6.1~$\times$~10$^{-4}$~m$^{-2}$s$^{-1}$ using a mean detection rate:
2.5~d$^{-1}$. \ensuremath{\beta}-meteoroids were not detected.
Clusters could be a closely-packed debris cloud or a particle breaking
up due to electrostatic fragmentation after high charging.

\end{abstract}

\section{Introduction}

A population of cosmic dust mixed with a population of man-made debris
exists within the Earth's magnetosphere. Measurements of these provide
the data samples for studies of the interplanetary dust particles that
travel through our magnetosphere from the outside and for studies of
the local byproducts of our space endeavors. Even though instruments
to detect natural meteoroids and space debris particles have been
flown in Low Earth Orbits (LEO) and on interplanetary missions, very
little information on the particle environment for Earth orbits above
about 600 km altitude have been obtained. In particular, knowledge
about debris particles smaller than 0.5~-~1~m in the geostationary
(GEO) region was largely unknown before GORID.

\section{Review of the GORID detector and its data}

The GORID impact detector is the refurbished engineering model of the
Ulysses dust detector, which detects particles by impact ionization
methods \citep{Drols:98,Drols:2001}. In this method of detection, a
particle impacting the detector at hypervelocity speed creates a
plasma of electrons and ions. The electrons and ions generated during
the impact are measured separately; the electrons are collected at the
target (\textit{Q}e), and the ions are collected at the ion collector
(\textit{Q}i). A few ions are further intensified and measured by
a channeltron (\textit{Q}c) behind the main ion collector grid. The negative
(electron) and positive (ion) charges generated upon impact range from
10$^{-16}$~C to 10$^{-8}$~C. The charge on the particle itself, as it
enters the detector, can be measured by the charge grids \textit{Q}p.

The velocity and mass of the impacting particle are deduced from the
rise-time and total intensity of the measured plasma signals using
empirical calibration curves. The rise-times of the measured plasma
signals are independent of the particle mass, and decrease with
increasing particle speed. Given the sensitivity and the calibration
of the instrument, GORID can detect particles with a mass down to
10$^{-17}$~kg.

The instrument is a shallow cylinder with an entrance
aperture of 43~cm and hemispherical target of area 0.1~m$^2$ and a 
viewing angle of 140\degg. The
GORID detector was in a geostationary location at 80\degg\ East
longitude until June~2000, when the satellite was moved to 103\degg\
East. Detailed information on the instrument design and data handling is
given in \cite{Gru:92b}, and a full description of the instrument
calibration and events classification is given in \cite{Gru:94a}.

In the absence of any anomalies, the procedure to determine speeds and
masses from GORID data is potentially straightforward: Laboratory
calibration is applied to convert digital signals to: electron, ion
and channeltron charges (\textit{Q}e,\textit{Q}i,\textit{Q}c) particle speeds derived
from target and ion collector rise-times (\textit{v}e, \textit{v}i). Debris
particles are discriminated (statistically) by impact speed and timing
of events (time of day, year, clustering). Constraints on individual
derived orbits (based on speed and instrument pointing direction) can
also be used. In general, determining the mass is via 
\vspace{-.25cm}
$$Q = k m^{\alpha} v^{\beta} , \quad \eqno(1) $$ 

\noindent where $k$, ${\alpha}$, and ${\beta}$  are derived from
laboratory calibration.  

Extensive calibrations were performed for the twins of GORID: the
Ulysses and Galileo dust detectors. Additional calibration tests for
GORID confirmed that the established calibration was still applicable.
The GORID signal amplitude calibration converts from digitized values
of the signal amplitudes to amplitudes of charge \textit{Q}e,
\textit{Q}i, \textit{Q}c and \textit{Q}p for the GORID dust detector. 
The calibration of
the relationship between the speed of the impactor and the rise-time
of the signals from the target and ion grid was performed using the
Heidelberg Electrostatic Accelerator. The mass of an impacting
particle can be derived from the charge to mass ratio for the ion or
electron charge, as a function of velocity with a linear (power law)
relationship: ${\rm log}(Q_i/m) = -1.063 + 3.375  \; {\rm log} (v) $ using
calibration data supplied by E. Gr{\"u}n (MPI-K Heidelberg). The
velocity calibration for the GORID dust detector converts from
digitized values of rise-times for the target and ion grid signals to
impactor impact speed (\textit{v}e , \textit{v}i).

Reliable impact speeds are the most important discriminator between
interplanetary dust and debris. Accurate speeds are required for mass
determination from impact charges because of the strong dependence on
speed of the charge to mass ratio (Eqn.~1). Unfortunately, there is
almost no correlation between the individual impact speeds derived from
the ion and electron rise-times. The reasons for this poor correlation
are not known, but could be due to a combination of the following: 1)
noise can cause apparent slow rise-times, 2) high particle charges can
also produce longer rise-times, 3) impacts on the sidewall of the
detector \cite{McDonnell:2000}. If we apply a single speed to all
particles (e.g. a typical weighted mean interplanetary dust impact speed
for GEO of 30 km s$^-1$ it can introduce a potential error of up to
10$^5$ in mass for debris particles that in reality may have impact
speeds as low as 1 km s$^-1$. This is due to the high value (3.4) of the
velocity exponent $\beta$ in Eqn.~1 from which the masses are derived.

Therefore the typical procedure to determine the impactor particle
speeds, and hence, masses, could not be implemented. If we could find
other characteristics of the measurements for separating debris
from the interplanetary (IP) dust, then mean speeds can be assigned
and hence fluxes. One characteristic of the measurements is the 
charge.

\section{Distribution of \textit{Q}{\small p} Charges}

The GORID data contains many particles with apparent high charge.
Table~\ref{Qpcharges} illustrates the distribution of particle grid
charges, \textit{Q}p, for the highest quality GORID events. We divide the
GORIDÕs charge grid detections (\textit{Q}p) into five \textit{Q}p classes
(labelled A to E) which demonstrate peculiar charge values (high,
negative) on the dust/debris particles.

\begin{table}[htpb]
\caption [Distribution of \textit{Q}p charges]{Distribution of \textit{Q}p charges}
\begin{tabular}{|llll|}\hline
{ \small Group}  & \textit{Q}p  &  {\small Number} & {\small Implied Properties}  \rule{0in}{2ex} \\[1ex]\hline\hline
{\small A} & {\small 4$\times$10$^{-10}$}  & {\small 463 }& {\small Saturated } \rule{0in}{2ex}\\
{\small B}  & {\small 4$\times$10$^{-10}$ -- 10$^{-12}$ } & {\small 1191} & {\small High -$v_e$ charge}   \\
{\small C} & {\small 10$^{-12}$ -- 10$^{-13}$ }   & {\small 1223 } &  {\small Medium -$v_e$ charge }\\
{\small D} & {\small 10$^{-13}$--0} & {\small 205 } & {\small Medium -$v_e$ charge} \\
{\small E}   & {\small $>$~0} &  {\small 230} & {\small +$v_e$ charge}   \\
\hline
\end{tabular}  
\label{Qpcharges}
\end{table}

Since potentials are not expected to exceed 10~V \cite{Mukai:2001},
then we have either very large grains, (10~$\mu$m--10~cm), or the
grain charging mechanism is not fully understood, or the 	\textit{Q}p values
are unreliable. We explored the \textit{Q}p values further by checking the
correlation for between electron and ion signals for -\textit{v}e and
+\textit{v}e charges. The \textit{Q}p data appear noisy with no
systematic biases, therefore, we decided to use \textit{Q}p as diagnostic
property of charging. The individual \textit{Q}p values may have
large uncertainties, but it is possible that the 
charges can provide diagnostic information for distinguishing IP and
debris populations, especially if coupled with another property
of the impacts, temporal variations. In Sect.~\ref{chproc}, we identify mechanisms
that could produce high negative equilibrium potentials on short 
timescales. Next we describe the temporal variations of GORID's impacts.

\section{Temporal Variations}

Temporal variations (clustering, diurnal, seasonal) may provide
additional constraints for statistical separation of populations.

\subsection{Clustering}

Very large variations in daily event rates led to the identification
of clusters of events, some of which re-occurred on consecutive days
at the same local time \citep{Drols:98,Sved:2000}. They were
interpreted as clouds of aluminium oxide debris resulting from the
firing of solid rocket motors. IP dust particles,
including \ensuremath{\beta}-meteoroids, have distributions in space
which are much larger than the scale of the GORID detector. Their
arrival times are therefore expected to have a random distribution
with a mean dictated by the particle flux for the particular pointing
geometry of the detector. Clustering of events could occur if an IP
particle fragments very shortly before impact, an extremely low
probability event. Clustering of events is therefore indicative of a
debris source and may provide a selection criteria for statistical
separation of the debris and IP populations.

\begin{figure}[thpb]
\centering
\includegraphics[width=0.95\linewidth]{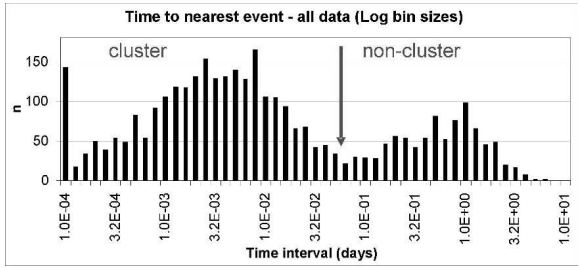}
\caption{\rm Distribution of time intervals between particle detections
used to determine cluster and no-cluster membership.}
\label{detCluster}
\end{figure}

An event is defined to be a cluster member if the time interval to the
closest event is less than a ``clustering limit''. The mean rate for
the highest-quality impact events is 1.83 day$^{-1}$, corresponding to
an interval of 0.55~d. The distribution of times between events (see
Fig.~\ref{detCluster}) shows a bimodal distribution with one component peaking at
about the expected ``random'' rate and the other with very much
shorter times, indicating clustering. The limit of cluster membership
is therefore defined at 0.05~days. IP particles would be expected to
show approximately random time intervals (with a slightly wider than
Gaussian spread due to the expected diurnal asymmetry). Typical time
intervals between non-cluster events are 0.2 -- 2 days and for cluster
events, seconds to \ensuremath{\sim}1 hour. 74\% of events are in
clusters.

The bimodal distribution of time intervals between events is 
apparent for all negative charge \textit{Q}p classes. \textit{Q}p 
class E data (positively charged particles) have no component 
of clustered events. Both the charge sign and the random nature 
of the detection times are consistent with \textit{Q}p, class E events 
being entirely interplanetary in origin.

\subsection{Diurnal and Seasonal Variation}

\begin{figure}[thpb]
\centering
\includegraphics[width=0.95\linewidth]{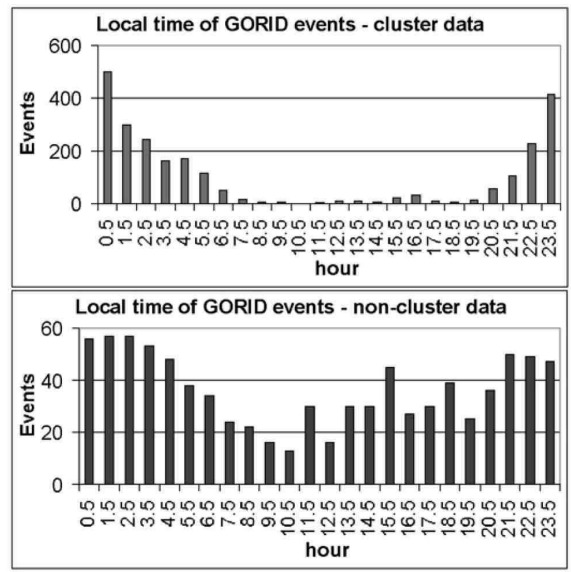}
\caption{\rm Distribution of impact detections as a function of time of
day for cluster (upper panel) and non-cluster (lower panel) data.}
\label{LTE}
\end{figure}

Cluster events (Fig.~\ref{LTE}) show very strong daily asymmetry whereas 
non-cluster data show a factor \ensuremath{\sim}2.5 asymmetry with the peak 
around midnight 
local time, entirely consistent with an IP origin.
Debris are concentrated near midnight local time except during 
summer. In early summer, clustered events are concentrated near 
5 am, just at the time when \ensuremath{\beta}-meteoroids may be expected 
to be detected. However, there is no evidence for an enhancement 
in the non-clustered data at this time suggesting an alternative 
explanation for these events.

\subsection{Crossing of Equatorial Plane in Magnetotail}   

The GORID pointing geometry implies preferential detection of clusters, e.g.
debris, after recent crossing of the equatorial plane in the magnetotail. 
This may be because debris are physically constrained to this 
region or because some process occurring in this region makes 
them detectable.

\section{On the highly-charged particles}   \label{chproc}

Several competing mechanisms  at 1~AU charge a debris or IP dust
particle: photoelectric effect, electron and ion collection, and
secondary electron emission. IP particles are expected to be
positively charged (appearing in the 	\textit{Q}p channel, E
class). The situation for orbital debris is more complex. Field
emission prevents further charging of a dust particle once it
possesses a sufficiently strong surface potential.

If one assumes we have extremely highly-charged particles, then we can
look at the dust charging conditions in GEO which might cause such an
effect. Previous work \cite{McDonnell:2000}\ studied in depth the
charging of dust/debris particles in Earth orbit from 2~R$_{\rm E}$ to
15~R$_{\rm E}$ (R$_{\rm E}$~=~Earth radius = 6.378~$\times$~10$^6$~m).
It considered two ranges of magnetospheric activity: solar ``typical"
(or quiet) and solar ``active" (or disturbed). The study used plasma
data for typical earth plasma conditions, defined by \textit{K}p=1,
\textit{A}p=120, and plasma data for active  Earth plasma conditions,
defined by \textit{K}p=5, \textit{A}p=1200.
The study found that the charging currents of dust/debris particles
are poised in a delicate balance, so that changes in the plasma
parameters, for example between quiet and disturbed plasma conditions,
and changes in particle material properties can have large effects on
the particles' charges. The next four cases (A,B,C,D) illustrate the
variations of these charging effects. More details can be found in
~\cite{McDonnell:2000}.

The next example illustrates conditions between quiet and disturbed
plasmas when the material properties, as indicated in the next box,
for charging the particle, remain the same. The equilibrium potential
charges were calculated assuming currents for ion and electron
collection from the plasma, photoelectric emission and secondary
electron  emission in a plasma with number densities and energies
given in Table~\ref{changingplasm}.

\vspace{-.3cm}

\begin{center}
\framebox{\parbox[h]{3.0in}{ $\delta_\textrm{m}$=1.4, $E_\textrm{m}$=180eV,
$\chi$=0.1, $a$=1\mic, $\rho$=2.3g/\pccm, $T_\textrm{photo}$=2.0eV
}}
\end{center}

Here, the particle material properties: $\delta_\textrm{m}$,
E$_\textrm{m}$, $\chi$, $a$, $\rho$, and $T_\textrm{photo}$ are the
number of secondary electrons (yield) at a characteristic (maximum)
energy at which the release of secondary electron peaks, the
photoelectric constant (1 = fully conducting, 0.1 = fully dielectric),
the radius and density of the  particle, and the energy of the
spectrum of the released photoelectrons in a Maxwellian distribution,
respectively. Table~\ref{changingplasm} gives the changing plasma
parameters, first, Case A, for a quiet plasma at 0 local time (Earth's
shadow) in GEO, and second, Case B, for disturbed plasma at 0 local
time in GEO.

\begin{table}[h]
\begin{minipage}{2.6in}
\renewcommand{\footnoterule}{}
\renewcommand{\baselinestretch}{0.8}
\begin{center}
\caption[Plasma Parameters for Cases A \& B]
{Plasma Parameters for Cases A \& B} \label{changingplasm}
\renewcommand{\baselinestretch}{1.5}
\begin{tabular}{|lllll|}
\hline
{\small Plasma} & { $n_\textrm{e}$ (1)} & { $n_\textrm{i}$ (2)} & {  $kT_\textrm{e}$ (3)} & { $kT_\textrm{i}$ (4)} \rule{0in}{2ex}\\   
{\small Type} & {\small (cm$^{-3}$)} & {\small (cm$^{-3}$)} & {\small (eV)} & {\small (eV)}  \rule{0in}{2ex}\\
{\small Quiet} & {\small 1.0 / 0.3} & {\small 0.6 / 0.8} & {\small 450 / 4000} & {\small 12 / 9000} \rule{0in}{4ex}\\
{\small Disturbed} & {\small 2.0 / 4.0} & {\small 2.0 / 2.3} & {\small 170 / 2400} & {\small 40 / 11000}\\
\hline  
\multicolumn{5}{l}{\footnotesize (1) Number densities for (hot/cold) electrons, (2) for (hot/cold) ions.}\rule{0in}{2ex}\\
\multicolumn{5}{l}{\footnotesize (3) Energies for (hot/cold) electrons, (4) for (hot/cold) ions. }\\
\end{tabular}
\end{center}
\end{minipage}
\end{table}

The resulting equilibrium potentials are {$U_\textrm{pot}$=2.2V} (Case
A) and {$U_\textrm{pot}$~={-2136.2}V} (Case B). The top charging
process is the electron collection current. The latter equipotential
current is high enough to be in the field emission
range~\cite{McDonnell:2000}. One plausible reason for the high
(theoretical) charge is multiple roots for equilibrium potential
\citep{Mey:82}, which are especially important in plasmas with high
temperatures or densities. For example, our numerical experiments that
increased only the electron density in the plasma pushed the
equilibrium potential to the next root, resulting in large negative
potentials, when we increased the electron number density two and a
half times ~\cite{McDonnell:2000}.

The charging of the debris/dust particles is particularly sensitive to
the secondary electron emission currents, so that a particle with
surface properties of a low yield material can change from having an
equilibrium potential a few Volts positive to an extremely highly
negative potential, assuming energetic or highly dense plasma
conditions which did not change, as illustrated in the next two cases:
C and D and Table~\ref{case3}. The material properties for the
charging calculation are indicated in the boxes. Notice that the only
changes between the two cases are the secondary electron yield (from
2.4 to 1.4) and the maximum energy for that secondary electron yield
(from 400 to 180~eV).

\begin{center}
\framebox{\parbox[h]{3.0in}{ $\delta_\textrm{m}$=2.4, $E_\textrm{m}$=400eV,
$\chi$=0.1, $a$=1\mic, $\rho$=2.3g/\pccm, $T_\textrm{photo}$=2.0eV,
\,\,{$U_\textrm{pot}$=3.7V} \,\,\textit{Case C}}}
\end{center}

\begin{center}
\framebox{\parbox[h]{3.0in}{ $\delta_\textrm{m}$=1.4, $E_\textrm{m}$=180eV,
$\chi$=0.1, $a$=1\mic, $\rho$=2.3g/\pccm, $T_\textrm{photo}$=2.0eV,
\,\,{$U_\textrm{pot}$=-3034V} \,\,\textit{Case D}}}
\end{center}

\vspace{-.25cm}

\begin{table}[h]
\begin{minipage}{2.5in}
\renewcommand{\footnoterule}{}
\renewcommand{\baselinestretch}{0.8}
\begin{center}
\caption[Plasma Parameters for Cases C \& D]
{Plasma Parameters for Cases C \& D}\label{case3}
\renewcommand{\baselinestretch}{1.5}
\begin{tabular}{|llll|}
\hline
$n_\textrm{e}$ (1) & $n_\textrm{i}$ (2) & $kT_\textrm{e}$ (3) & $kT_\textrm{i}$ (4)  \rule{0in}{2ex}\\
(cm$^{-3}$) & (cm$^{-3}$) & (eV) & (eV)  \rule{0in}{2ex}\\
{2.0 / 1.0} & {2.0 / 1.0} & {300 / 7000} & {300 / 7000} \rule{0in}{4ex}\\
\hline 
\multicolumn{4}{l}{{\footnotesize (1) Number densities for (hot/cold) electrons, (2) for (hot/cold) ions.}}\rule{0in}{2ex}\\
\multicolumn{4}{l}{{\footnotesize (3) Energies for (hot/cold) electrons, (4) for (hot/cold) ions. }}\\
\end{tabular}
\end{center}
\end{minipage}
\end{table}


\section{Derived Fluxes and Mass Distributions}

In the absence of any reliable speed determination, mean speeds 
are assigned for debris (clustered) and IP (non-clustered) events. 
Fluxes are then determined by multiplying the speeds
with the number of particles in a unit time and integrating 
over the GORID detecting surface. 

Representative impact speeds for IP particles and space debris 
are assigned for non-clustered and clustered events respectively.
For IP particles, we used an IP model \cite{McBride:2000}, which
gives a speed distribution weighted by 
the impact plasma detector response \textit{v}$^{3.4}$, with which
we find a mean speed 
of 31.4~km s$^{-1}$. For debris particles, we used the following.
Particles in near-GEO orbits will have impact speeds of a few 
hundred meters per second. However, at these low speeds, they 
are not likely to produce large amounts of impact plasma. The 
dominant contribution (i.e. highest relative speeds) of debris 
are expected from impactors in geo-transfer orbits (GTO).
The weighted (by response/geometry/source) mean speed~=~2.6~km~s$^{-1}$. 
For accessible retrograde GTO orbits, \textit{v}~=~4.5~km s$^{-1,}$ and 
for circular retrograde GEO orbits \textit{v}$_{max}$~=~6.1~km~s$^{-1}$.

The GORID raw data files gave the number of particles per unit time.
The data from the GORID experiment is distributed in three different 
sets of files: science, count, and housekeeping files. Details
of the data format can be found in: \cite{McDonnell:2004}.
The total number of the highest quality events in the science file is 3349 
(2477 clustered, 872 non-clustered) in 1827 days. In the count 
file data there are \ensuremath{\sim}5486 events. We can use the known 
properties of the science data to assign the count file events 
as either IP or debris before calculating the fluxes.

After the counts are determined from the data files, they
were corrected for ``dead time". 
Interplanetary events, which are detected essentially at random, 
suffer a ``dead time'', equivalent to the clustering time of 0.05 
days. The total dead time for IP (i.e. non-clustered) detections is 
0.05~$\times$~3349~=~167~d, so that the dead time correction 
is~=~1827/(1827-167) = 1.101. Therefore, the total expected number 
of interplanetary events is 872~$\times$~1.101~=~960.
The detection rate of interplanetary events from the science data file 
is 960/1827~=~0.53~d$^{-1}$. The dead-time interval is slightly 
higher due to events in the count files, which exceed the maximum
event number of the count buffer of the instrument. After correcting for 
the 28 additional events, the total number of IP events is 988.
The total detection time is 1827~d giving a mean interplanetary flux 
of 0.54~d$^{-1}$.

The space debris flux is therefore calculated from all remaining 
events detected, i.e. 5486 -- 988 = 4498 in 182~d. The mean debris 
detection rate is therefore 2.46 d$^{-1}$.

\begin{figure}[thpb]
\centering
\includegraphics[width=0.95\linewidth]{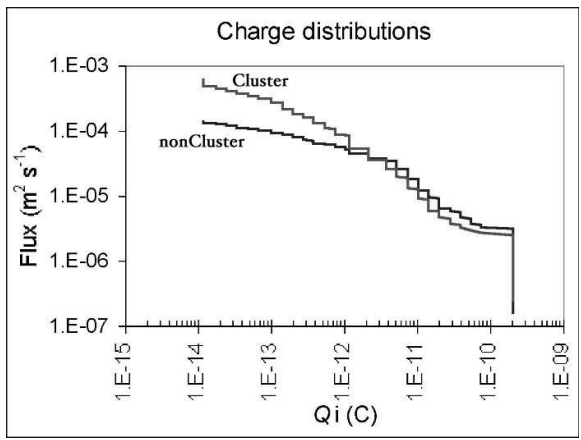}
\caption{\rm Cumulative fluxes on GORID as a function of ion grid charge.}
\label{detCumcharge}
\end{figure}

Cumulative fluxes are calculated for an equivalent flat plate 
detector pointing in the same direction as GORID, which has a 
detector area \textit{A}~=~0.1~m$^{2}$ and effective viewing angle of 
\ensuremath{\Omega}~=~1.45~sr: 
$F\left( {\succ Q_i} \right)=\left( {N\left( {\succ Q_i} \right)\;f\;\left( 
{\pi /\Omega } \right)} \right)\;/\;\left( {T\;A} \right)$, 
where $N\left( {\succ Q_i} \right)$ is the number of events with ion grid 
charge greater than \textit{Q}i, \textit{f} is a scaling factor to account for missing data (due to 
dead time and count file data, and \textit{T} is the time interval over which the observations were made.
Figure~\ref{detCumcharge}\ shows the calculated cumulative fluxes for cluster 
and non-cluster data. The ion grid signals are saturated at
 \textit{Q}i~=~2~$\times$~10$^{-10}$~C. The debris (cluster) data show relatively greater 
fluxes at smaller charges.

\begin{figure}[thp]
\centering
\includegraphics[width=0.95\linewidth]{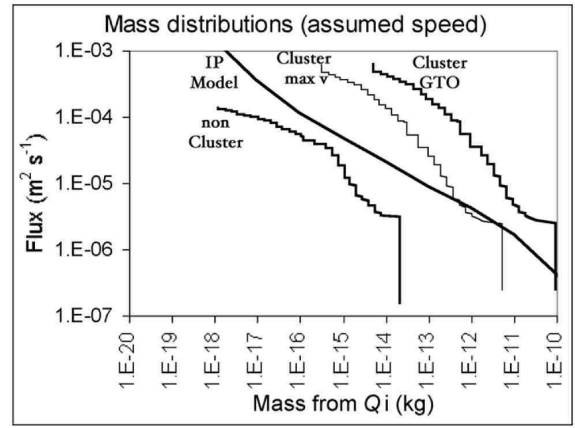}
\caption{\rm Cumulative mass distributions derived using a mean impact
speeds for interplanetary dust of 31 km s$^-1$ applied to non-cluster
data and either a mean (2.6~km~s$^-1$) or maximum (4.5~km~s$^-1$) GTO
impact speed for cluster data. An interplanetary dust model prediction
\cite{McBride:2000} is shown for comparison.}
\label{detCumMass}
\end{figure}

Figure~\ref{detCumMass}\ shows the cumulative mass distributions
derived using representative single values of impact speed for debris
(cluster) and IP (non-cluster) events compared with the interplanetary
model described in  \cite{McBride:2000}. The calculated interplanetary
flux in GEO is lower than the model by a factor of between 2 and 10
even though this model provides a good fit to impact cratering data
measured on spacecraft in LEO. The assumption of a single velocity is
important. The distribution of speeds of interplanetary dust ranges
from a few to 70 km s$^{-1}$ and the speed distribution used in the
model, derived from meteor data, may not precisely apply to the
smaller particles detected by GORID. A factor of 2 error in speed
results in a factor of 10 error in mass because of the v$^{3.4}$ power
relationship between impact speed and derived mass for a given impact
charge (Eqn.~1). The flux is defined by the measured impact rate. A
higher impact speed would result in lower masses for the entire
distribution and hence a lower flux at any specific mass (the actual
error in flux would depend on the mass distribution but would also be
a factor of 10 for a mass distribution index of -1). In addition, the
data have been converted from GORID detections to fluxes seen on a
flat plate detector pointing in the same direction. Debris fluxes are
calculated for the nominal mean impact speed for GTO of 2.6~km~s$^-1$
and for the maximum possible GTO impact speed of 4.5~km~s$^-1$. In
both cases, much higher fluxes are observed for debris than
interplanetary particles in this size range. Debris fluxes appear to
be \ensuremath{\sim}20 times the interplanetary flux model for masses
below 10$^{-14}$ kg.

We can test if the fluxes derived using the impact speeds of GTO
particles are reasonable by estimating the total mass of dust in the GTO
population. If we make a simple assumption of uniform spatial density of
debris in a spherical cloud from altitudes of 1000~km to GEO and a
particle mass of 10$^{-14}$ kg, we derive a total mass of such particles
around the Earth of a few hundred kg. This is likely to be an
overestimate because particles spend a large fraction of their orbital
time near apogee and the orbital distribution of the presumed sources
have non-uniform inclination distributions and apsides close to the
equatorial plane. Alternatively, some of the debris particles may be
moving significantly faster than the assumed speed, and hence be much
smaller, their large charge causing acceleration in the magnetosphere.
Without reliable impact speeds it is not possible to determine the true
fluxes to high precision. Uncertainties in mass of a least a factor of
10 are inevitable.

\section{Summary and Future Work}

Although the speeds derived from rise-times in GORID data are 
not reliable, the clustering of events can be used to discriminate 
between space debris and interplanetary particles. 
The detection rate of interplanetary events is 0.54 d$^{-1}$ and 
debris events is 2.46 d$^{-1}$.
The mean fluxes are 1.35~$\times$~10$^{-4}$~m$^{-2}$~s$^{-1}$ for IP and 
6.1~$\times$~10$^{-4}$~m$^{-2}$~s$^{-1}$ for debris at the detection 
threshold of \textit{Q}i~=~1.3~$\times$~10$^{-13}$~C.
The fluxes of interplanetary particles are reasonably close to 
the well-defined model prediction, allowing for the impact speed 
uncertainty. Detectability of debris may be influenced 
by charging mechanisms in the magnetosphere.

One possible explanation for the clusters could be a closely-packed debris
cloud or a slag particle breaking up under electrostatic fragmentation.
Work by \cite{Bunte:2004} demonstrated that a debris cloud from a GEO
insertion burn could stay together for weeks and months to give
multiple impacts with the same time stamp. A different work by Felix
van der Sommen (personal communication) showed that a closely-packed
debris cloud originating near GEO could be detected by GORID as a cluster, 
if: 1) initial
velocities are very low, and 2) they are charged enough.   
Another possibility is electrostatic fragmentation, which 
has been considered for clustered phenomena in the Earth's
magnetosphere before (e.g.~\cite{Fechtig:79}). Slag particles
from a rocket GTO or GEO burn might fragment upon entering the highly
energetic environment of the magnetospheric plasma sheet. These
possibilities need to be investigated further.

\vspace{1.5cm}
{\bf Acknowledgments}

We thank Eberhard Gr\"un (MPI-K, Heidelberg), Antal Juhasz (KFKI, Budapest), and
Doug Hamilton (University of Maryland), who played a significant role
in the earlier ESA study~\cite{McDonnell:2000}, leading to this work. ALG also thanks 
the anonymous referee for his helpful comments and the INAF
Istituto di Fisica dello Spazio Interplanetario for its financial
support.

\vspace{.5cm}


\bibliographystyle{esa_ref}
\bibliography{amthesis}

\begin{thebibliography}{10}

\bibitem{Drols:98}
Drolshagen, G., Svedhem, H., Gr\"un, E., Grafodatsky, O., and Prokopiev, U.
\newblock Microparticles in the geostationary orbit {GORID} experiment.
\newblock {\em COSPAR}, 1998.

\bibitem{Drols:2001}
Drolshagen, G., Svedhem, H., and Gr\"un, E.
\newblock Measurements of cosmic dust and micro-debris with the {GORID} impact
  detector in {GEO}.
\newblock {\em Proc. Third European Conference on Space Debris ESOC}, ESA
  SP-473, 2001.

\bibitem{Gru:92b}
Gr\"un, E., Fechtig, H., Giese, R.~H., Kissel, J., Linkert, D., Maas, D., and
  McDonnell, J.
\newblock The {U}lysses dust experiment.
\newblock {\em Astron. Astrophys. Suppl. Ser.}, 92:411--423, 1992.

\bibitem{Gru:94a}
Gr{\"u}n, E., Baguhl, M., Hamilton, D., Kissel, J., Linkert, D., Linkert, G.,
  and Riemann, R.
\newblock Reduction of {G}alileo and {U}lysses dust data.
\newblock Technical report, Max-{P}lanck-{I}nstitut f{\"u}r {K}ernphysik, 1994.

\bibitem{McDonnell:2000}
McDonnell, J.
\newblock Update of statistical meteoroid/debris models for {GEO}, final report
  of {ESA} contract 13145/98/nl/wk.
\newblock Technical report, European Space Agency, July 2000.

\bibitem{Mukai:2001}
Mukai, T., Blum, J., Nakamura, A.~M., Johnson, R., and Havnes, O.
\newblock {\em Interplanetary Dust}, chapter Physical processes on
  interplanetary dust, pages 445--507.
\newblock Springer Verlag, Heidelberg, Germany, 2001.

\bibitem{Sved:2000}
Svedhem, H., Drolshagen, G., Gr{\"u}n, E., Grafodatsky, O., and Prokopiev, U.
\newblock New results from in-situ measurements of cosmic dust Ð data from the
  {GORID} experiment.
\newblock {\em Adv. Space Res.}, 289:309--314, 2000.

\bibitem{Mey:82}
{Meyer-Vernet}, M.
\newblock {`Flip-flop'} of electric potential of dust grains in space.
\newblock {\em Astr. Ap.}, 105:98--106, 1982.

\bibitem{McBride:2000}
McBride, N. and Hamilton, D.
\newblock Meteoroids at high altitudes, final report of {ESA} contract
  13145/98/nl/wk.
\newblock Technical report, European Space Agency, July 2000.

\bibitem{McDonnell:2004}
McDonnell, J.
\newblock Processing, analysis and interpretation of data from impact
  detectors, final report of {ESA} contract 16272/02/nl/ec.
\newblock Technical report, European Space Agency, July 2004.

\bibitem{Bunte:2004}
Bunte, K.
\newblock The detectability of debris particle clouds, for final report of
  {ESA} contract 16272/02/nl/ec.
\newblock Technical report, European Space Agency, July 2004.

\bibitem{Fechtig:79}
{Fechtig}, H., Gr\"un, E., and {Morfill}, G.
\newblock Micrometeoroids within ten earth radii.
\newblock {\em Planetary and Space Science}, 27:511--531, April 1979.

\end{thebibliography}

\end{document}